# Computational Aeroacoustics of a Generic Side View Mirror using Stress Blended Eddy Simulation

K.K Chode [1], H. Viswanathan [2], K. Chow [3]

[1] *Sheffield Hallam University, Sheffield, United Kingdom, k.k.chode@shu.ac.uk.*
[2] *Sheffield Hallam University, Sheffield, United Kingdom, h.viswanathan@shu.ac.uk*
[3] *HORIBA MIRA Ltd, Warwickshire, United Kingdom, kevin-chow@horiba-mira.com*

**ABSTRACT:**
This paper presents a numerical investigation of aerodynamic noise generated by a generic side-view mirror mounted on a flat plate using the Stress Blended Eddy Simulation (*SBES*) coupled with the Ffowcs Williams and Hawkings (FW-H) equation. A grid evaluation study was performed using a standardised side-view mirror with a Reynolds Number ($R_e$) of 5.2 x$10^5$ based on the diameter of the model. The predictions for hydrodynamic pressure fluctuations on the mirror, the window and the sound emitted at various microphone locations are in good agreement with previously published experimental data. In addition, our numerical results indicate that yawing the mirror closer to the side window results in the flow being attached to the rear of the mirror resulting in an overall reduction in Sound Pressure Level (SPL) at several receiver locations.

## 1.INTRODUCTION

The noise generated near the side windows due to external fixtures such as side-view mirrors and its impact on passenger comfort is crucial for the automotive industry. Höld et al., (1999) developed a generic side-view mirror based on a half-cylinder and a quarter sphere to study the side-view mirrors in a more standardised way. Höld et al., (1999), and Siegert et al., (1999) predicted the noise generated by a generic side view mirror using experimental and Unsteady Reynolds Averaged Navier-Stokes (*URANS*) numerical approaches. Ask and Davidson (2005, 2009) conducted a comparison study between Detached Eddy Simulation (*DES*) and Large Eddy Simulation (*LES*) using both structured and unstructured meshes. The LES approach predicted a wide range of turbulent scales when compared to the DES model. The hydrodynamic pressure fluctuations are in good agreement, mainly at higher frequencies with LES. The far-field noise was computed using Ffowcs-Williams and Hawkings (*FW-H*) acoustic analogy. Yao et al., (2018) has compared both the DES and LES approaches using two different flow conditions. Their results indicated that the compressible LES approach resolved more fluctuations in the boundary layer than DES approach. Although several studies have been performed with standard mirror cases and with realistic mirrors (Dawi et al. (2018, 2019)), the effect of yawing the mirrors has not been investigated till date.

The commonly used RANS approaches are based on Eddy viscosity model, which are widely used in evaluating flow around bluff bodies (Read and Viswanathan (2020), and Guilmineau et al. (2011)).With recent advancement in turbulence modelling, Stress-Blended Eddy Simulation (*SBES*) (Menter (2018)) has shown to reasonably predict a smooth but rapid transition from RANS to LES in the Separating Shear Layer (*SSL*), and accurately predict the aerodynamic characteristics of standard geometry (Chode et al., 2020). However, the coupling capabilities of SBES model with acoustic analogies such as FW-H needs a thorough investigation. Therefore, the present study aims to investigate the noise generated by a generic side-view mirror using SBES with the Wall-Adapting Local Eddy-viscosity (*WALE*) model. The far-field noise generated by a side-view mirror is computed using the FW-H analogy. The authors will be validating the results obtained with experimental results by Ask and Davidson (2009). Further, the influence of yaw on the noise generated by a side-view mirror is investigated.

## 2. METHODOLOGY

Geometrically, the generic side-view mirror has a diameter ($D$) of 0.2 m and is positioned on a flat plate as shown in Fig.1 a). The computational domain is extended from the centre of the mirror to 15D upstream and 30D downstream. In the normal and spanwise direction, the boundaries are extended to 15D form the centre. The dimensions of the computational domain are consistent with previously simulated cases (Ask and Davidson (2005,2009)).

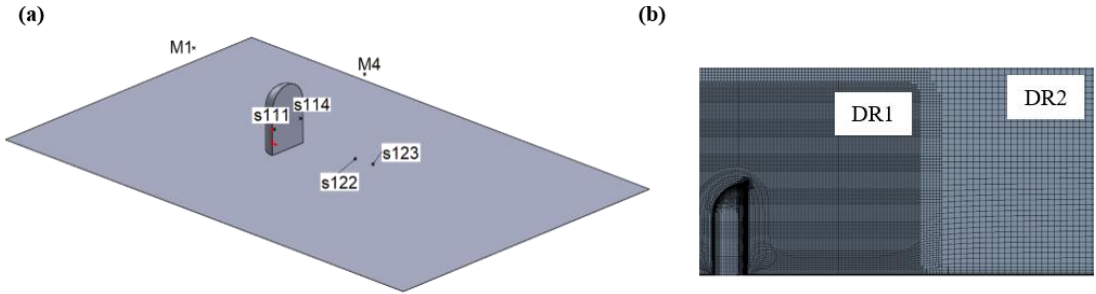

**Figure 1**. **a)** Sensor layout on Mirror and Plate, **b)** Overview of the Grid-3 Mesh on the XY plane.

Three meshes were evaluated in the present study and were generated using the Cut-Cell methodology that consists of unstructured hexahedral elements. The details of the mesh sizes used for the mirror are detailed in Table 1. For the Grid-3, the mesh size on DR1 seen in Fig.1 b) corresponds to 18 Points Per Wave, which results in resolving the frequency up to a range of 4kHz - 5kHz. The flow was considered to be incompressible with a Reynolds Number of $R_{eD} = 5.2 \times 10^5$ based on the mirror diameter with a constant inlet velocity ($U$) of 39 ms$^{-1}$. A constant pressure outlet was imposed on the outlet with symmetry on the extended boundaries. The no-slip condition was imposed on both the mirror and the plate while a slip condition was assigned to the ground. The flow was initialised with the solution obtained through steady-state RANS results from $k-\omega\ SST$ model. The time step used in the current study was $3 \times 10^{-5}$s, which corresponds to a Courant-Friedrichs-Lewy (*CFL*) number < 1. Both time-averaged quantities, pressure fluctuations and FW-H coupling were activated after simulating for ~ 60 convective cycles (60 x *D/U*).

**Table 1.** Comparison of mesh resolution used in the current study

|  | $\Delta x^+$ | $\Delta y^+$ | $\Delta z^+$ | PPW | **Cell Count** |
|---|---|---|---|---|---|
| **Grid – 1** (No Refinement) | 360 – 680 | < 1 | 360 – 680 | 5 | $1.1 \times 10^6$ |
| **Grid – 2** (One Refinement – DR1) | 240 – 520 | < 1 | 240 – 520 | 9 | $2.4 \times 10^6$ |
| **Grid – 3** (Two Refinements – DR1, DR2) | 120 – 520<br>120 – 320 (Edges) | < 1 | 120 – 520<br>120 – 320 (Edges) | 18 | $4.9 \times 10^6$ |

## 3. RESULTS AND DISCUSSION

### 3.1 Grid Evaluation Study

The average pressure coefficient predicted by Grid-3 is in good agreement with the other numerical results published using LES and DES turbulence model (see Fig.2 a)). The drag coefficient predicted by SBES is within the values of drag reported by Ask and Davidson (2009) and Chen and Li (2019) numerical simulations, as shown in Table 2. The

Hydrodynamic Pressure Fluctuations (*HPF*) predicted on the mirror and plate were plotted at four different locations as seen in Fig.1 a). All the grids investigated were able to capture the trend predicted by experimental data accurately up to the sensors cut-off frequency. As seen in Figs2 b)-e), an average of 11 dB difference can be observed in the maximum value predicted by the numerical results when compared against experiment ((Ask and Davidson (2009)). Thus, Grid-3 mesh settings are used to evaluate the effects of yaw on sound generation by the side-view mirror.

**Table2**
Comparison of Drag Coefficient against previously published numerical results

| Published cases | $C_D$ | Current Case | $C_D$ |
| --- | --- | --- | --- |
| Ask and Davidson (2009) – LES | 0.4497 | Grid-1 – SBES (WALE) | 0.4660 |
| Chen and Li (2019) – DES | 0.4895 | Grid-2 – SBES (WALE) | 0.4623 |
| Chen and Li (2019) – DDES | 0.4785 | Grid-3 – SBES (WALE) | 0.4600 |

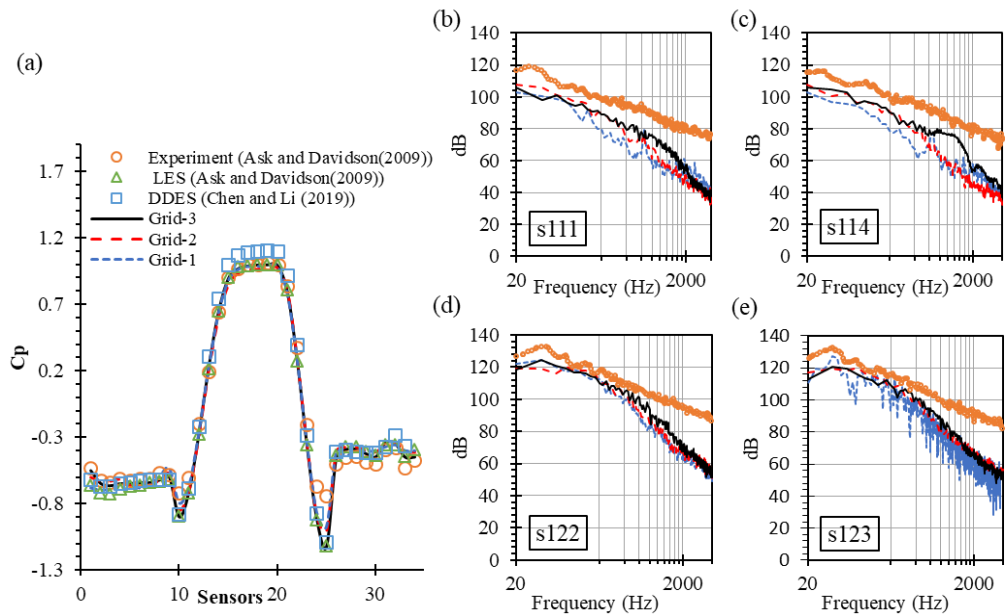

**Figure 2. a)** Average Cp over the mirror, SPL generated at **b)** s111, **c)** s114, **d)** s122, **e)** s123

### 3.2 Effect of Yaw on Sound Generation
The mirror is yawed by 4° such that the edge of the mirror is closer to the window. The yawing resulted in an overall decrease in the sound levels reported at most of the observer locations around the mirror (only M1 is shown in Fig. 3g) expected for observer M4 as shown in Fig .3h). This increase in sound level at M4 is due to the location of the observer is located close to the high fluctuating pressure region over the plate, as shown in Fig 3c). At this location (highlighted in black in Fig 3), there is a 6.1% increase in the overall pressure fluctuations on the plate due to yawing the mirror. Further, HPF reported in the wake of the mirror at s122 and s123 also indicate a decrease in maximum dB reported which is due to the shrinkage of the separation bubble as highlighted in red Fig 3 b) and d).

### 4. CONCLUSION

Computational Aero-acoustics of a generic standard mirror using the SBES turbulence model coupled with FW-H, has been investigated. A grid evaluation study was conducted to arrive at a mesh that is capable of reasonably predicting the pressure fluctuations on the plate exerted by the flow. The SPL trend predicted by numerical results shows a good agreement with experiment up to the local cut off frequency, and an average drop of 9.2% in dB is seen at the maximum HPF levels when compared to the experimental results. A reduction in overall SPL is predicted at specific observer locations when the mirror is yawed by 4°, although increases in SPL occur more locally at some observer location around the mirror. Further investigations will be carried out to determine the effect of several yaw angles and add-on features, such as rain-gutters on the flow and their influence on both the HPF and the overall SPL levels at various microphone locations.

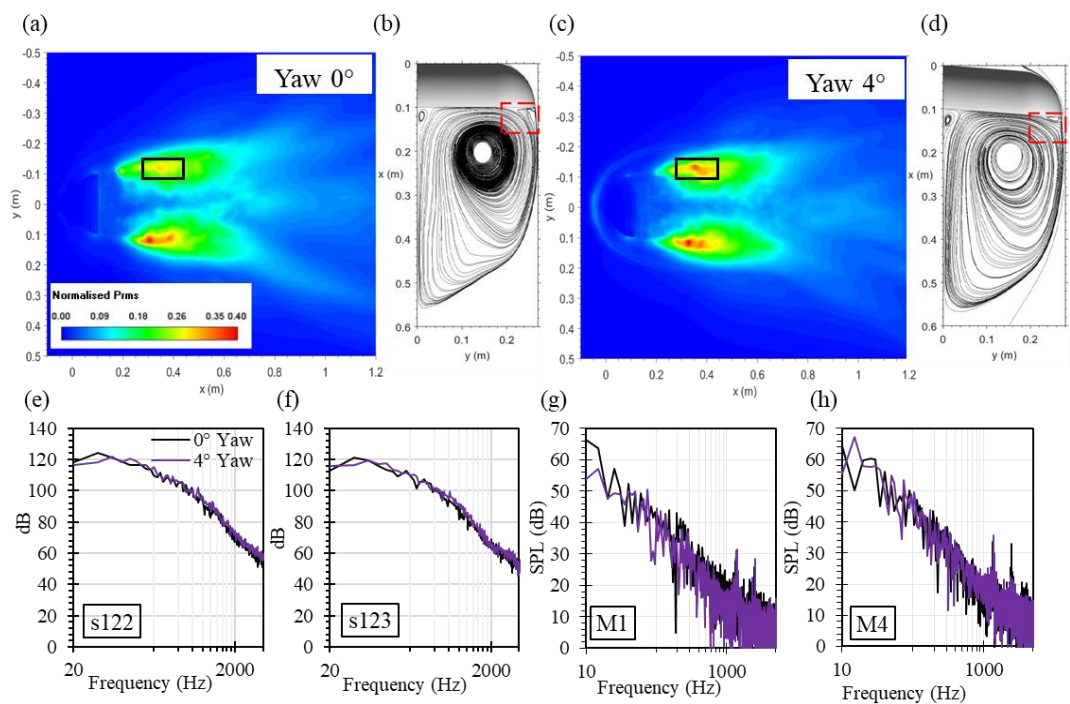

**Figure 3.** Time-averaged Pressure Fluctuations on a yz plane at y = 0.02m **(a & c)**, streamlines of the wake behind the mirror **(b & d),** HPF at s122 **(e)**, s123 **(f)** and SPL at the observer at M1 **(g)** and M4 **(h)**.

**ACKNOWLEDGEMENTS**
The authors would like to thank ANSYS for the Academic Research Partnership grant. This work is supported by the Vice-Chancellor PhD scholarship between Sheffield Hallam University and HORIBA MIRA Ltd.